\documentstyle[12pt,aasms4,psfig]{article}   
\slugcomment{January 22, 2001}
\lefthead{Garc\'ia-Barreto et al.}
\righthead{NGC 3367}

\begin{document} 
 
\title{Fabry Perot H$\alpha$ observations of the Barred Spiral NGC 3367}

\author{J.Antonio Garc\'{\i}a-Barreto and M. Rosado}
\affil{Instituto de Astronom\'{\i}a, Universidad Nacional Aut\'onoma de
M\'exico, Apartado Postal 70-264, M\'exico D.F. 04510 M\'exico}

\begin{abstract}

We report the gross properties of the velocity field of the barred spiral galaxy NGC 3367. We obtained the following values : inclination with respect to the plane of the sky, $i=30$; position angle (P.A.) of receding semi major axis, P.A.$=51^{\circ}$ and systemic velocity, V$_{sys}=3032$ km s$^{-1}$. Large velocity dispersion are observed of up to 120 km s$^{-1}$ in the nuclear region, of up to 70 km s$^{-1}$ near the eastern bright sources just beyond the edge of the stellar bar where three spiral arms seem to start and in the western bright sources at about 10 kpc. Deviations from normal circular velocities are observed from all the disk but mainly from the semi circle formed by the string of south western H$\alpha$ sources. An estimate of the dynamical mass is M$_{dyn}=2\times10^{11}$ M$_{\odot}$.  

\end{abstract}

\section{Introduction}

NGC~3367 is an SBc barred spiral galaxy considered isolated at a distance of 43.6 Mpc behind the Leo Spur group of galaxies (i.e. using H$_0=75$ km s$^{-1}$ Mpc$^{-1}$  \cite{tul88}), with a distant neighbor NGC 3419 at $900\pm100$ kpc away. At a distance of 43.6 Mpc,  1$''$ corresponds to 210 pc. Its optical appearance in the broadband filter I ($\lambda=8040$ \AA~ from \cite{gar96a,gar96b}) is shown in Figure 1; it suggests a galaxy almost face on but compilation values from the published papers indicate that its inclination with respect to the plane of the sky is not well determined. Although the weak optical emission originates from an almost axisymmetric disk, the bright emission shows an asymmetry such that only the south western region looks symmetric half disk, but not the north eastern region. The stellar bar has a radius of only $16''$ (3.3 kpc) oriented at P.A. $63^{\circ}\pm5^{\circ}$, and there is an optical structure consisting of several H$\alpha$ knots resembling a ``bow shock'' at a radius of about 10 kpc from the nucleus from south east to north west (\cite{gar96a,gar96b}).  VLA observations at 4$''$.5 angular resolution at 1.46 GHz show radio continuum emission from two lobes extending to at least 6 kpc straddling the nucleus, most likely off the plane, as well as weaker emission from the disk (\cite{gar98}). Soft X-ray emission has been reported by \cite{gio90}.Single dish 21 cm HI observations indicate values of a systemic velocity of v=3035$\pm8$ km s$^{-1}$ and a total hydrogen mass of M(HI)=$5\pm2\times10^9$ M$_{\odot}$ (\cite{hel81,huc85,mir88,sta88}). Optical systemic velocity values are v=2850$\pm50$ km s$^{-1}$ (\cite{ver86,dev91,ho95,ver97,dev91,ho95,ho97}).

\begin{figure}[t]
\psfig{figure=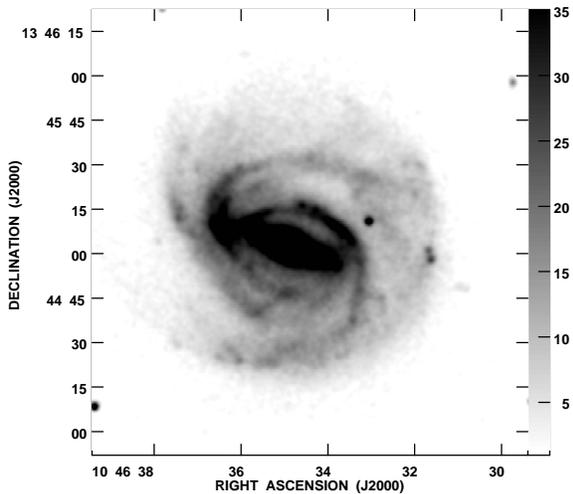,width=8cm,angle=-90}
\caption[garcia3367fp.fig1.ps]{
Optical continuum emission, from NGC 3367, in the broadband filter I ($\lambda 8040$ \AA~) convolved to an output beam full width at half maximum of $1''.5$ (dictated from the seeing conditions, after \cite{gar96b}). Greys scale is proportional to peak brightness distribution in arbitrary units.}  \label{fig1}\
\end{figure}

In this paper we present new Fabry Perot H$\alpha$ observations of NGC~3367 carried out with the PUMA equipment attached to the San Pedro M\'artir 2.1m optical telescope in Baja California, M\'exico. The velocity field thus obtained indicates that differential rotation appears to dominate. Departures from near circular rotation are                                                                                                                                                                                                                                                                                                                                                                                           clear from the central region and also from the outer bright semi circle to the south west. The fitted rotation curve shows a slow rise and flat part. $\S$ 2 presents the observations and calibration. $\S$ 3 presents the data reduction and analysis. $\S$ 4 presents the velocity field. $\S$ 5 presents the discussion and finally $\S$ 6 presents the conclusions.

\section{Observations and Calibration}

Two dimensional velocity field for the ionized gas in NGC 3367 was obtained by using the UNAM Scanning Fabry-Perot (FP) interferometer PUMA on 1999 February 8 \& 9. This instrument is currently in use at the f/7.9 Ritchey-Chretien focus of the 2.1 m telescope at the Observatorio Astron\'omico Nacional at San Pedro M\'artir, B.C., M\'exico. 

The PUMA setup consists of a scanning Fabry-Perot interferometer,
a focal reducer 2:1 which brings the initial focal ratio from f/7.9 to f/3.95, a f/3.95 camera, a filter wheel, a calibration system
and a Tektronix CCD detector of 1024$\times$1024 pixels (Rosado et al.
1995). Since NGC 3367 has angular dimensions less than $3'$ we used only a central region of the CCD of $512\times512$ pixels. In this way, we observed a field of view of 5$'\times5'$ with a pixel size of $0''.6\pm0''.01$. 

To obtain FP data cubes of NGC 3367, we used an interference filter
with a central wavelength of 6620 \AA ~and a narrow band (30 \AA) in order to isolate the redshifted H$\alpha$ emission of this galaxy. The gap between the plates of the FP interferometer,with an interference order of 330 at  H$\alpha$,  was scanned over a free spectral range and a  total of 48 interference channels (each 120 seconds of integration time) were obtained with a velocity increment of 19 km s$^{-1}$. Thus, the data cubes are of 512$\times512\times$48.

Calibration cubes were obtained  before and after the cubes corresponding to NGC 3367. The line at 6598.95 \AA ~of a neon lamp was used for wavelength calibration. 

The raw data were calibrated in wavelength using the specialized software CIGALE developed at the Marseille Observatory (Le Coarer et al. 1993) and object cubes were produced consisting of 48 velocity maps of 512$\times512$ with final image scale of 0$''$.6 pixel$^{-1}$ and a spectral sampling of 19 km s$^{-1}$. No attempt was made for absolute calibration of the H$\alpha$ emission.

\section{Data Reduction and Analysis}

The cube was exported from CIGALE in a fits format to the AIPS package in order to continue with the data analysis. Preliminary inspection of the emission at different velocities showed that a bright south west star was elongated in a position angle (P.A.) 13$^{\circ}$. Having no idea whether it is a double star from our observations or from POSS plates nor anything arising from instrumental effect that one could possibly correct for, it was decided to convolve the 48 planes to an output gaussian beam of 6$''.5~\times4''.1$ at P.A. 13$^{\circ}$, based on the fits on the sw star, with the task CONVL. With this beam the experiment was aimed mainly on the velocity field with a medium angular resolution. This, as mentioned later, turned out to be crucial in not being capable to study the kinematics of the central most region of the galaxy. For the best continuum image it was decided to align as best as possible all the image planes. For this purpose, careful position determination using IMSTAT and IMFIT were used in AIPS for the nuclear position and a bright south west star in the field. It was decided to obtain shifts with respect to plane 25 which showed the maximum brightness (in arbitrary units) for both the nuclear region and the star. Shifts less than 3.6 pixels in RA and 0.9 pixels in DEC were found for the nuclear region and less than 1.5 pixels in RA and 0.63 pixels in DEC were found in the south west star. Individual planes were corrected for the corresponding shifts found using the program LGEOM in AIPS. A continuum offset was determined, from regions outside galaxy emission, and corrected for each plane with the program COMB. In order to obtain a continuum emission, from the cube, to be subtracted from each plane the data was displayed for every velocity channel and it was visually decided which channels had no emission. Finally channels 2 to 13 and 37 to 48 were chosen. First an image was produced from channels 2 to 13 using the program SQASH and another image using the channels 37 to 48. Both images were then averaged using the program COMB in order to have a final continuum image. This continuum image was subtracted from the cube. 
Finally, the program MOMNT in AIPS was used in order to obtain the moments zero (integrated intensity), one (velocity) and two (velocity dispersion), after smoothing the data with a box car of size 5 and 7 in velocity and space coordinates. After careful inspection of the results using different cutoff values for flux and intensity it was decided to allow only data using fluxes and intensities above values that guaranty to determine a velocity field where weak, but above noise level, emission was detected. Figure 2 shows the moment 0 or total integrated intensity map of the H$\alpha$ emission. The map is similar to the previous map obtained with a direct image camera taken at the same observatory (\cite{gar96a,gar96b,gar98}). Figure 3 shows the channel maps with heliocentric radial velocities indicated corresponding to channels 18 to 33. Figure 4 shows the isovelocity contours plotted on a grey representation of the same H$\alpha$ distribution shown in Fig. 2. Figure 5 shows velocity dispersion and Figure 6 shows the residual velocity field (observed minus model).

\begin{figure}[t]
\psfig{figure=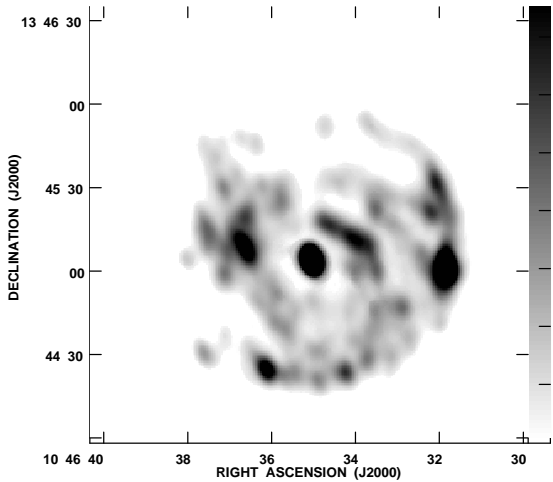,width=8cm,angle=-90}
\caption[figure=garcia3367fp.fig2.ps]{
Integrated intensity of H$\alpha$ emission, moment 0 map in NGC 3367. Contours and greys scale are proportional to peak brightness distribution in arbitrary units.}  
\label{fig2}
\end{figure}

\begin{figure}[t]
\psfig{figure=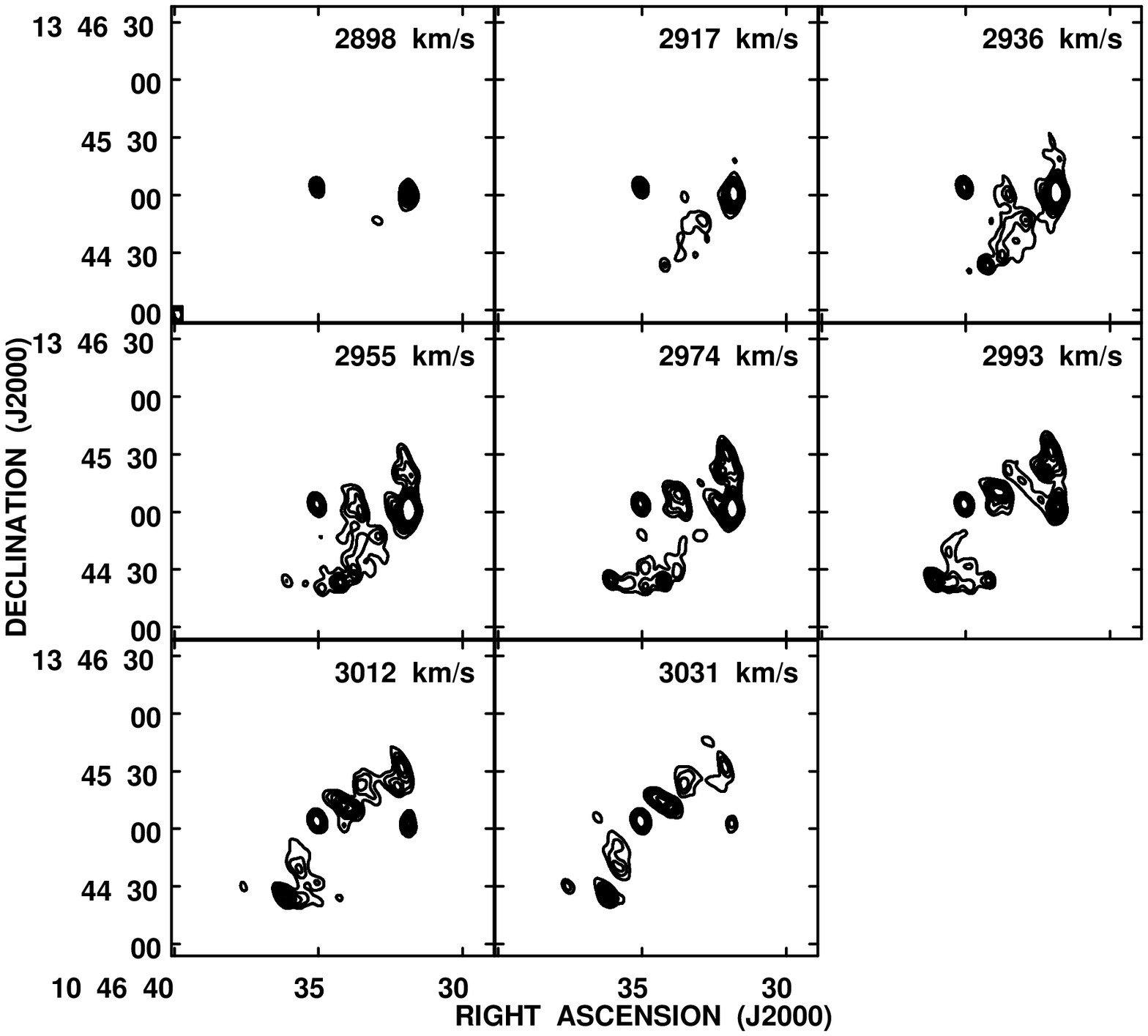,width=8cm,angle=0}
\psfig{figure=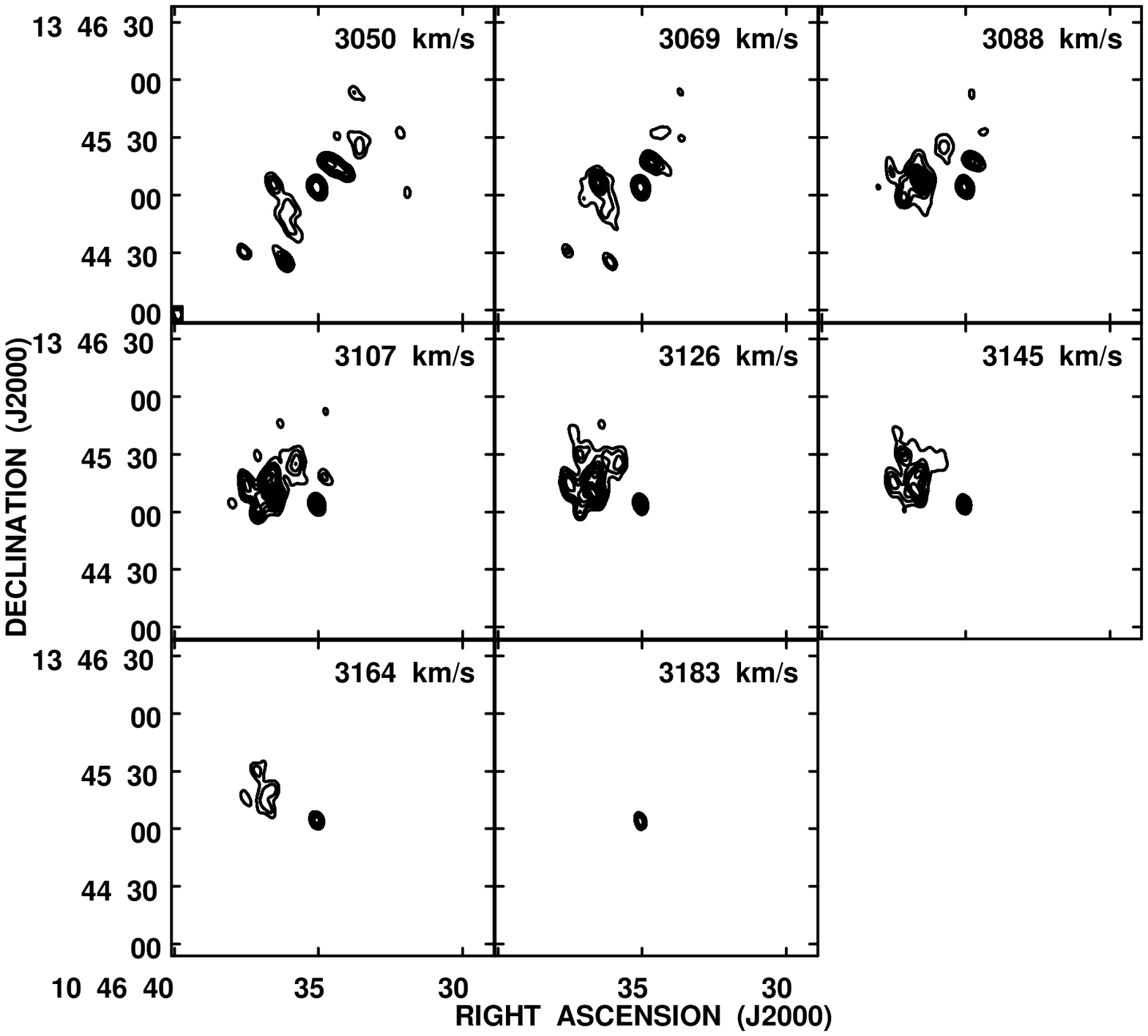,width=8cm,angle=0}
\caption[figure=garcia3367fp.fig3a.ps]{
Individual channel maps of intensity distribution in NGC 3367. Contours are proportional to peak brightness in arbitrary units. Velocities in upper right corner are heliocentric velocities and correspond to a chaneel separation of 19 km s$^{-1}$. These maps correspond to channels 18 to 25 in upper panel and channels 26 to 33 in lower panel.}  
\label{fig3}
\end{figure}

The program GAL in AIPS was used in order to get the rotation curve. The observed velocity at a given radius is given by the expression

\begin{center} 
V$_{obs}=$ V$_{sys}~+~$ V$_{\theta}$(R,$\theta$)sin$i$cos$\theta~+~$ V$_R$(R,$\theta$)sin$i$sin$\theta~+~$ V$_z$(R,$\theta$)cos$i$, 
\end{center}

where $i$ is the inclination of the disk with respect to the plane of the sky, $\theta$ is the azimuthal angle in the plane of the galaxy, V$_{\theta}$(R) is the azimuthal velocity at a radius R, V$_R$(R) is the radial velocity component at a radius R, and V$_z$(R) is the vertical velocity component at a radius R, V$_{sys}$ is the systemic velocity of the galaxy (\cite{mih81}). The program GAL assumes, as a first approximation, that the gas is in near circular orbits in a plane, that is, it considers the V$_R$(R) and V$_z$(R) velocity components equal to zero. This assumption will be not valid in the case of NGC 3367 as it will be seen below. The orientation in space of a galaxy is described, in general, by four parameters: the position of the center (x,y), the position angle of the receding semi major axis (P.A.) and the inclination of the plane ($i$). Other parameters involve the systemic velocity, V$_{sys}$, the maximum rotation velocity, V$_{max}$, and the radius at which occurs the maximum rotation velocity, R$_{max}$ (\cite{moo85}). Our first attempt was to allow to vary all parameters x,y, P.A., $i$, V$_{sys}$, V$_{max}$, R$_{max}$ using both the receding and approaching sides. The fitted values for the different parameters were x=10$^h$ 46$^m$ 35$^s$.04, y=$+13^{\circ}$ 45$'$ 06$''.4$, P.A.$=52^{\circ}$, $i=30$, V$_{sys}=3034$ km s$^{-1}$, V$_{max}=208$ km s$^{-1}$ and R$_{max}=49''$. The fitted values for the x and y position corresponded to the kinematical center $\sim~0''.5$ west in RA and $\sim~2''.1$ north in Dec. from the center of surface brightness maximum. Varying the values of some parameters, while holding fixed others, we tried to get the best rotation curve.

\begin{figure}[t]
\psfig{figure=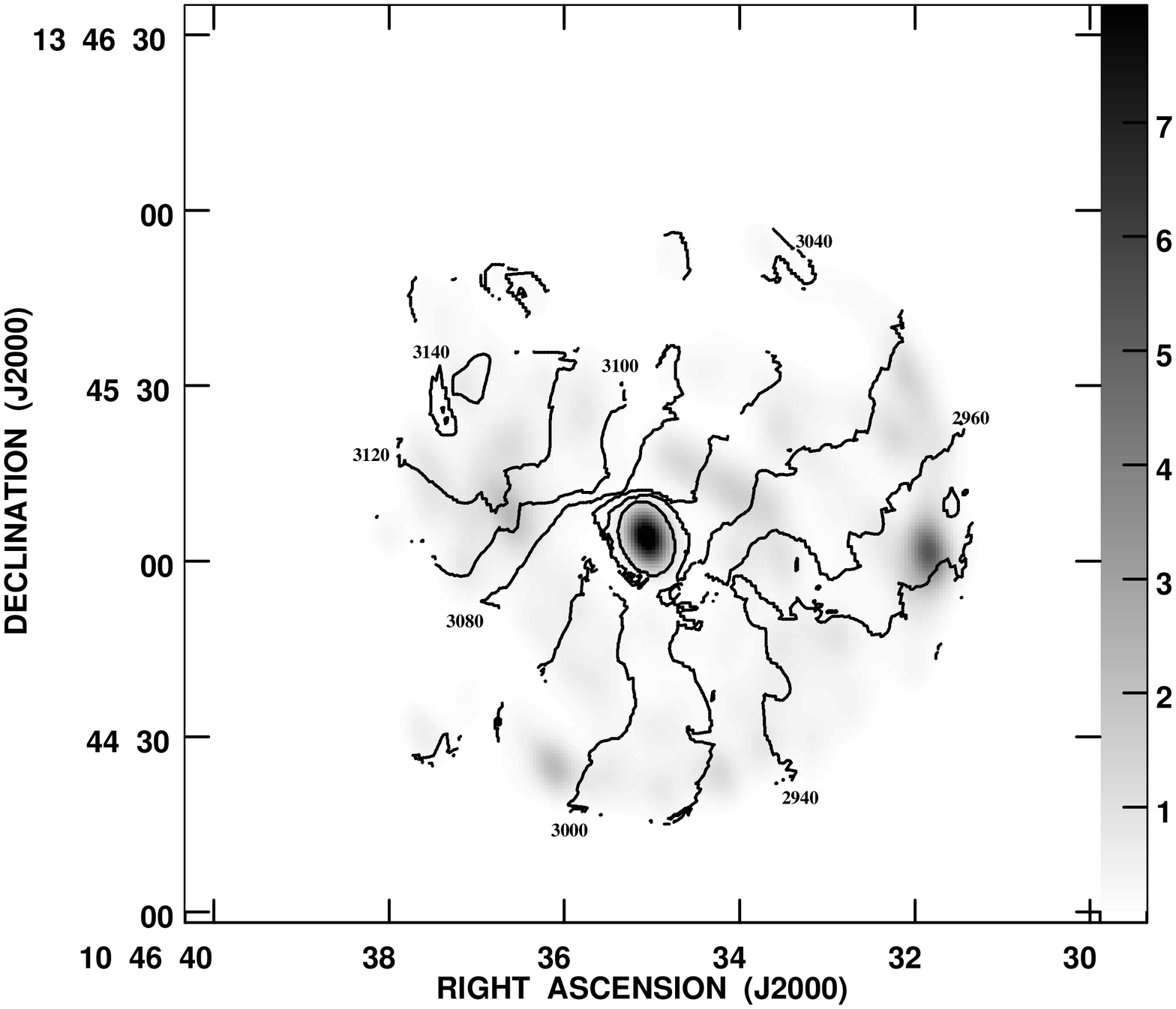,width=8cm,angle=-90}
\caption[figure=garcia3367fp.fig4.ps]{
Isovelocity contours (moment 1) superimposed on the total integrated intensity (moment 0) map. Kinematic receding semi major axis  was found to be at P.A.=51$^{\circ}$. Velocities shown are 2940, 2960, 3000, 3040, 3080, 3100, 3120 and 3140 km s$^{-1}$. V$_{sys}=3032$ km s$^{-1}$ }  
\label{fig4}
\end{figure}

\section{The Velocity Field}

	It is not advisable to make an elaborate many parameters fit to the rotation curve, however, the fitting of the orientation parameters is by far the most important task (\cite{moo85}). First it was noticed that the kinematical center fitted, with all parameters free to vary, was off by about $-0''.5$ in RA and +2$''.3$ in Dec from the center of the brightness distribution. At this point it is necessary to mention that the coordinate grid to the cube was anchored in such a way as to make coincide the maximum H$\alpha$ brightness distribution and the peak of emission in the high resolution VLA 3.6 cm radio continuum from the central region (which we think corresponds to the nucleus, \cite{gar00}). We decided to fix the position of the x,y coordinates to the maximum of the integrated intensity  distribution, that is, at $\alpha$(J2000)=10$^h46^m35^s.06$, $\delta$(J2000)=$+13^{\circ}45'04''.1$.
We ran the program fixing different parameters and fitting others as listed in Table 2. 

	Average values of V$_{max}$ for small diameter Sc galaxies with high inclination angles are between 150 km s$^{-1}$ and 200 km s$^{-1}$ (\cite{rub85,rub99}). Since the value of the inclination $i$ has been reported in the range from 6$^{\circ}$ to 37$^{\circ}$, a search for reported values of the inclination of NGC 3367 followed. Table 1 lists inclination angles found in the literature for NGC 3367 and the method used.

	In general, determination of the inclination of a galaxy has been done in any of three methods: i) From multiparametered fits to detailed high resolution radio synthesis or Fabry Perot maps; ii) from axial ratio corrected for spiral arm structure and iii) from axial ratio according to the formula (\cite{hub26,mih81})

$$cos^2i=\{(b/a)^2-(b_i/a_i)^2\} / \{1~-~(b_i/a_i)^2\}$$

where $a$ and $b$ are the apparent major and minor semi axis of a galaxy and $a_i$and $b_i$ are the intrinsic major and minor semi axis. In general for a flattened spheroid the intrinsic axial ratio is taken to be 0.2 (\cite{hol58}).

	As stated by Aaronson, Mould and Huchra (1980) method i) is the best method for measuring sin$~i$ since it is the only one that actually measures the inclination of the rotating gas and is based on quantitative analysis of kinematic structure, except for kinematic departures from differential rotation, as HI warps (\cite{bos81}). Aaronson, Mould and Huchra (1980) found that inclinations determined from axial ratios from RC2 were too face on by about 3$^{\circ}$.

	To evaluate our inclination and position angle of the receding semi major axis values we looked at the residual velocity field (observations minus model) for the different values of the inclination, namely, $i=20^{\circ}$, $i=25^{\circ}$ and $i=30^{\circ}$ and two values for the position angle of the receding semi major axis, P.A.$=46$ and P.A.=51$^{\circ}$. The rotation curve was slightly modified but mainly only the velocity scale. We did run the program using only $\pm$60$^{\circ}$ from the P.A. of the line of nodes of both the receding and approaching sides in order to look for asymmetry deviations but we did not find any and finally we also tried leaving  out the central 10$''$.
Finally, the residual velocity field with least asymetries was the one using an inclination of $i=30^{\circ}$. 
Values for inclination and maximum rotation velocity are listed in Table 2.

\begin{figure}[t]
\psfig{figure=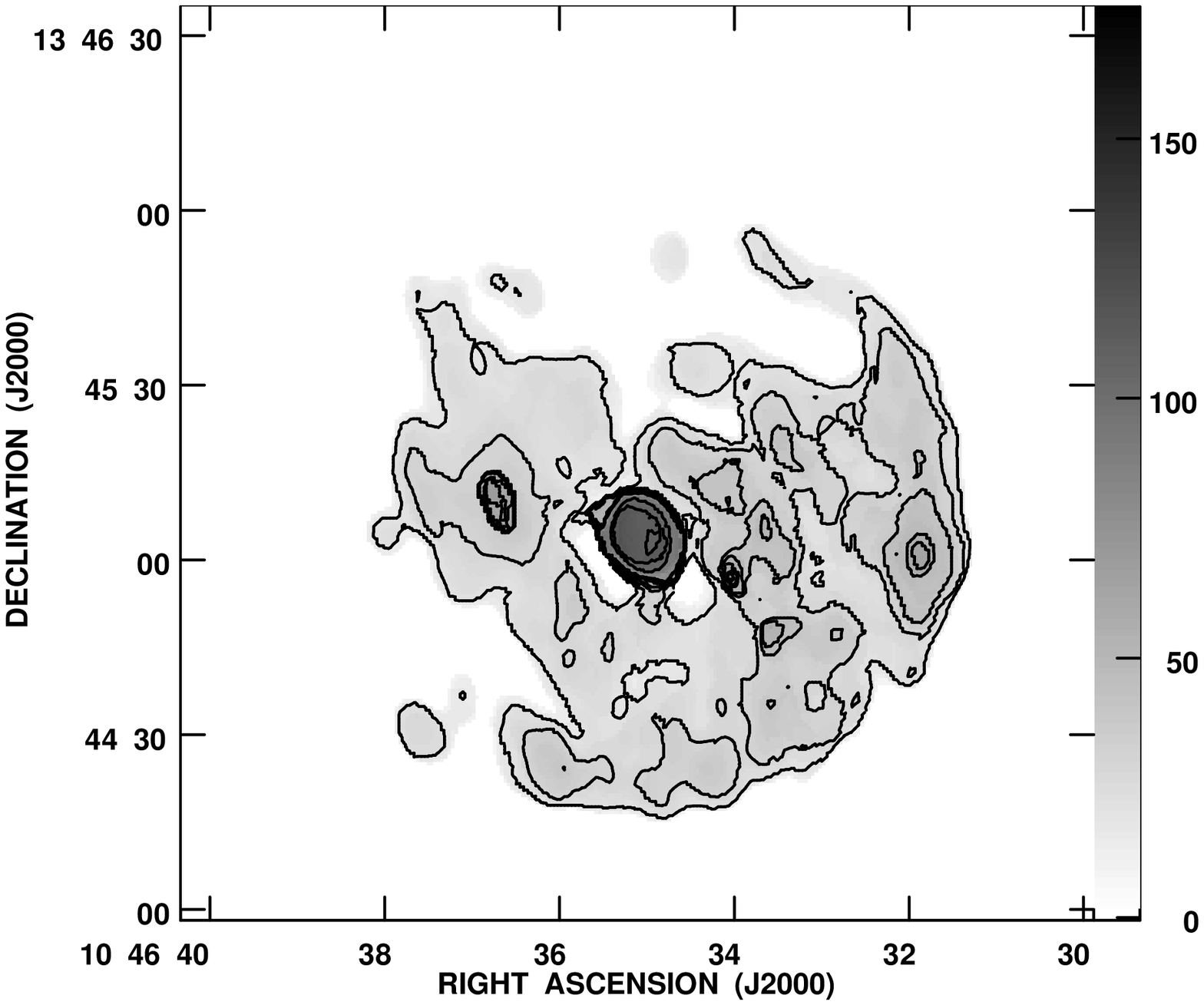,width=8cm,angle=-90}
\caption[figure=garcia3367fp.fig5.ps]{
 Dispersion velocities (moment 2) map in contours and greys scale. Velocity contours correspond to 20, 30, 40, 50, 60, 65, 70, 100 and 110 km s$^{-1}$}  
\label{fig5}
\end{figure}

The maximum velocity is in agreement with the estimated maximum velocity value obtained from HI measurements, as : V$_{max}$ = (1/2) ($\Delta$ V$_{HI}^{50}$)/(sin $i$), since for NGC 3367 $\Delta$V$_{HI}^{50}$ = 206 km s$^{-1}$ (\cite{huc85}). 
It is smaller, though, by about 50 km s$^{-1}$ from the estimated value if using the width of the HI line at 20 \% ($\Delta$V$_{HI}^{20}$). 

\begin{figure}[t]
\psfig{figure=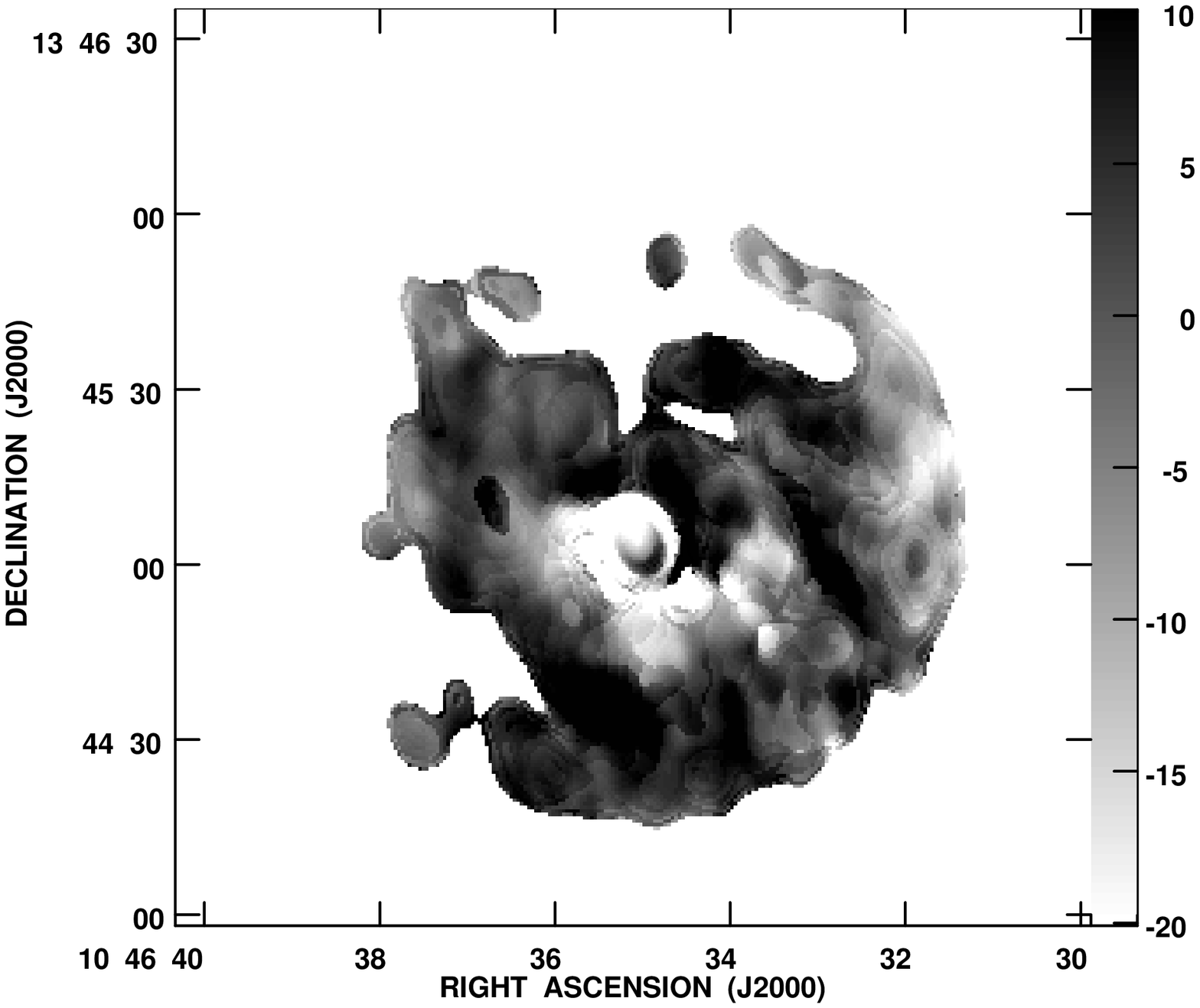,width=8cm,angle=-90}
\caption[figure=garcia3367fp.fig6.ps]{
Residual velocities (observed - model) map in grey scale. Scale shown corresponds to -20 up to +10 km s$^{-1}$. Notice : 1) a velocity asymmetry in the central region at a position angle (P.A. at about 18$^{\circ}$) compared to the P.A. of the disk at about 51$^{\circ}$, possibly suggesting a different inclination from the central region, and 2) the negative residual velocities mainly from the south west semicircle structure. This map was used in order to get the minimum residual velocities and asymmetries when fitting for the position angle of the semi major axis and the inclination of the galaxy.}  
\label{fig6}
\end{figure}

The final rotation curve, assuming as usual azimuthal symmetry and zero vertical and radial components of the velocity vector, is shown in Figure 7. 
This rotation curve can be fitted as a first approximation by a Brandt curve (\cite{bra60}) with V$_{max}=210$ km s$^{-1}$ at R$_{max}=52''$ (11 kpc). 
An estimate of the dynamical mass using V$_{max}$ and R$_{max}$ has been obtained. 
Radio interferometric detection of atomic hydrogen (from larger radii than the maximum distance where optical H$\alpha$ emission was detected) and a model to account for non-zero vertical or radial components are a must in
order to make a detailed modeling of the mass distribution (bulge, disk,
halo) (see for example \cite{ken87,van90,fre93,oll96,sic97,ryd98}).

	Finally  we obtained for NGC 3367, based on our Fabry Perot observations, the following values  :
\begin{enumerate}

\item Center at $\alpha$(J2000)=10$^h46^m35^s.06$, $\delta$(J2000)=$+13^{\circ}45'04''.1$
\item Position Angle of receding major axis, P.A.=51$^{\circ}\pm3^{\circ}$
\item Inclination with respect to plane of sky, $i=30^{\circ}, {+2^{\circ}}, {-5^{\circ}}$
\item Systemic Velocity, V$_{sys}=3032~\pm3$ km s$^{-1}$
\item Maximum Rotation Velocity, V$_{max}=210$ km s$^{-1} \pm15$ km s$^{-1}$
\item Radius at maximum velocity, R$_{max}=52''\pm3''$
\item Dynamical Mass inside 52$''$, M$_{dyn}=2^{+0.5}_{-0.25}\times10^{11}$ M$_{\odot}$

\end{enumerate}

\section{Discussion}

	NGC 3367 is known to exhibit two large ($\approx 2$ kpc diameter)  radio continuum lobes straddling the nucleus at a distance of about 6 kpc (\cite{gar98}); while at high resolution  there is a compact nuclear source ($\leq~70$ pc in diameter) surrounded by a structure at a radius of less than 350 pc (\cite{gar00}), with probable presence of WR stars from optical spectroscopy (\cite{ho95}). 
H$\alpha$ optical imaging observations, show an unresolved bright source, most likely associated with the compact nucleus and weak extended emission covering the innermost 5$''$ but the existence of a possible circumnuclear structure is difficult to confirm because of poor angular resolution (\cite{gar96a,gar96b}). Our Fabry Perot observations were not the exception, since we had also very low final angular resolution, as mentioned earlier. Spectra from NGC 3367 in the red optical region indicates a moderate width of the H$\alpha$ line only to be considered to have an HII nucleus (\cite{ho97}). Only two out of ten SBc galaxies from the list observed by Martin (1995) have bars' semi axes smaller than the bar in NGC 3367, suggesting that the bar in NGC 3367 is small given its Hubble type. From high resolution radio continuum observations the lobes are connected to the central region but this emission is most likely also out of the plane and not associated with any dust or shock regions along the stellar bar since these are expected to be on the leading side of the bar, south east and north west assuming that the bar is rotating counter clockwise as suggested by trailing arms (as seen in other galaxies, for example NGC 1365 [\cite{jor95}]). The presence of radio continuum lobes out from the nuclear region and X-ray emission indicate the presence of activity and suggest an inflow process to the very center of the galaxy. The high dispersion velocity found in the central region (of up to 120 km s$^{-1}$) might be associated with the circumnuclear structure and probable outflow (since emission is observed at different velocity channels, see Fig. 3). Starburst driven galaxy outflows have been reported for several strong far infrared emitters (\cite{hec90}). Faint broad optical emission lines have been modeled as associated with shells of interstellar material smoothly accelerated after breakout (\cite{ten97}). The existence of an unresolved radio continuum source ($\leq~70$ pc in diameter) hints that the source of energy is a compact nucleus; however optical spectroscopy fails to detect emission from high ionization atoms (ie. Fe) from NGC 3367 as detected from other active galactic nuclei (\cite{ho95}). The source of the central activity is still to be resolved and is beyond the scope of the present work. Since the raw rotation curve in Figure 7, most likely includes non zero vertical and radial components of the velocity at different radii, the derived angular velocity is therefore ambiguous as well as the radial frequencies (since $\kappa$(R) depends on $\Omega$ and d$\omega$/dR). However, if 
the south west ring were the result of an outer Lindblad resonance (OLR) at a distance between 8.5 kpc and 10 kpc, a possibility that was discarded because of lack of symmetry on the north east side (\cite{gar96a}), then the estimated constant angular velocity of the bar would be close to 43 km s$^{-1}$ kpc$^{-1}$ and co-rotation (CR) would be at a radius between 3.5 kpc and 4 kpc, which would correspond to about 1 and 1.25 the radius of the stellar bar. These estimates must clearly be taken with caution and only as a zero order, but it is fair to say that they are in agreement with values given for other barred galaxies (\cite{but96}). However, nothing could be said about the distance from the nucleus expected for an inner Lindblad Resonance (ILR) which in general for Sc galaxies is expected to be at larger radii than for Sa galaxies, simply as a result of mass concentration in early type galaxies. The structure in high resolution radio continuum observations between 0.3 kpc and 0.45 kpc from the compact nucleus remains to be proven if it lies near an inner Lindblad resonance (ILR).
 
\begin{figure}[t]
\psfig{figure=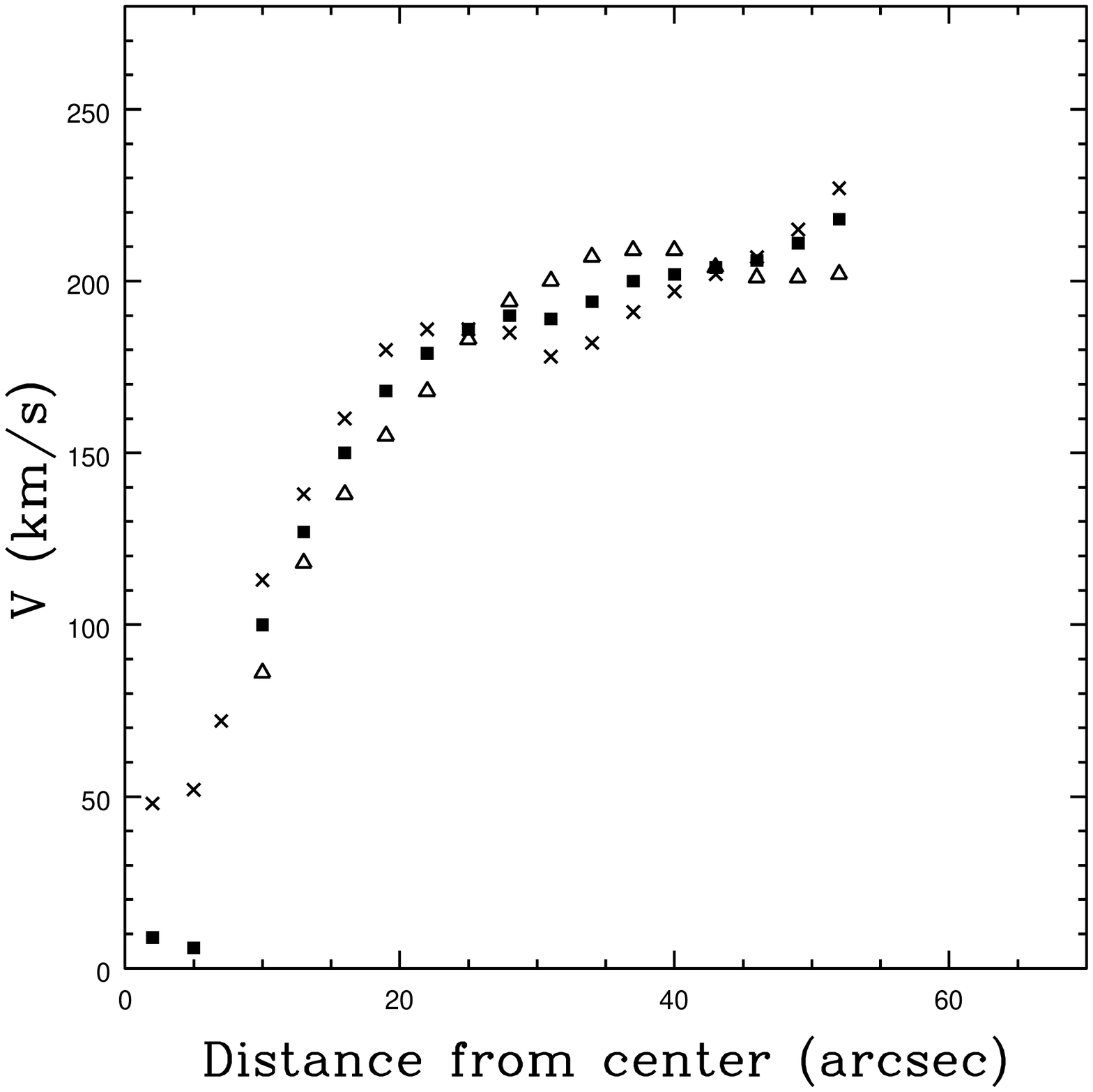,width=8cm,angle=0}
\caption[figure=garcia3367fp.fig7.ps]{
Final rotation curve for NGC 3367 determined from Fabry Perot observations using the receding side only (open triangles), the approaching side only (crosses) and both the receding and approaching sides (filled squares). Error uncertainities in velocity vary from 2 to 15 km s$^{-1}$. 1$''$ corresponds to 210 pc.}  
\label{fig7}
\end{figure}

	Deviations from circular orbits have been routinely observed in almost every spiral galaxy (\cite{teu86,gio88,rot90,jor95,sic97,bea99}). Substantial noncircular streaming motions and large residual velocities were found in the inner bar region of NGC 1365 (\cite{teu86,jor95}). In the case of NGC 3367, isovelocity contours from a) the central region, and b) from the south west (approaching side) outer most regions of the galaxy are observed to deviate from normal circular motion. For example, the isovelocity contours from the south in NGC 3367 indicate an observed bluer velocity of about 20 to 30 km s$^{-1}$, at a radius of about 8.5 kpc from the nucleus, than the expected value; a similar behaviour is observed from the western bright sources (see Fig. 4).  In particular from inspection of the channel velocity maps (Fig. 3), the western bright region shows from the first channel at 2898 km s$^{-1}$ up to the channel at velocity 3031 km s$^{-1}$ suggesting an expanding bubble of about 65 km s$^{-1}$ centered at a radial velocity of 2965 km s$^{-1}$ which would correspond to the expected circular velocity. The central and bright eastern regions show up in different velocity channels possibly suggesting the existence of expanding gas also, however more detailed observations are needed in order to confirm this idea.
 
Large velocity dispersions from the central region in NGC 3367 might be associated with an outflow. Reasons in favor are: a) a plasma outflow is definitely there since NGC 3367 presents two radio continuum lobes connected to the nucleus at a radii of 6 kpc, b) a probable circumnuclear structure (\cite{gar00}) with WR stars (\cite{ho95}) and c) X-ray emission (\cite{gio90}). 

A possible interpretation for large velocity dispersions from the south west semi-ring (see Fig. 4) might be that these sources are part of a structure with a higher radial velocity, from 10 to 40 km s$^{-1}$, than expected at their radii assuming circular orbits (\cite{bea99}). This radial velocity could be the result of coeval evolution of HII regions evolving in a normal disk or could be the result of a large scale phenomenon (ie bow shock (\cite{kri83}), outward wave (\cite{lyn76,too78}),etc.). Yet, another possibility is that the semi-ring of H$\alpha$ sources is at a higher inclination with respect to the plane of the sky and therefore inclined with respect to the galaxy interior disk. If the outer region were a warp (\cite{bin92}) it would show from atomic hydrogen observations. 

	Finally, Gr\"osbol (1985) has studied the mass distribution of many spiral galaxies from azimuthal intensity profiles based on POSS red plates, including NGC 3367. Values found by him are: radius with a mean surface brightness of 23.5 magnitudes/arcsec$^2$ r$_{23.5}=74''$; inner radius in which the mean intensity could be approximated by an exponential disk, r$_i=6''$;  outer radius in which the mean intensity could be approximated by an exponential disk, r$_o=46''$ and scale length of exponential disk, r$_l=24''.3$. Baggett, Baggett \& Anderson (1998) also have reported a bulge disk decomposition for 659 spiral galaxies including NGC 3367 based on major axis brightness profiles from the Photometric Atlas of Northern Bright Galaxies by Kodaira, Okamura \& Ichikawa (1990). They fitted de Vaucouleurs law profiles for bulges and a truncated exponential disk (after Kormendy). In particular for NGC 3367 they report a disk scale length of 8$''$, a disk truncation radius of 26$''.6$ and a bulge efective radius of 25$''.8$. It is pertinent here to repeat that NGC 3367 has a stellar bar in a P.A.=63$^{\circ}\pm5^{\circ}$ and a radius of only 16$''$ (\cite{gar96a,gar96b}), and thus the bulge disk decomposition by Baggett at al. includes the stellar bar inside what they consider a bulge. A detailed model of the mass distribution in NGC 3367 including bulge, disk and halo is necessary and it would benefit from not only optical surface photometry and velocity field but also from extended atomic hydrogen gas and velocity distributions.
	
\section {Conclusions}

We have carried out Fabry Perot H$\alpha$ emission observations from the 
barred galaxy NGC~3367 field in the disk of the galaxy with the PUMA equipment at the San Pedro M\'artir 2.12m telescope in order to determine the velocity field in NGC 3367. Importat results can be summarized as follows: i) We have determined the values for different parameters as follows: inclination of the disk with respect to the plane of the sky to be $i=30^{\circ}$; position angle of the receding semi major axis, P.A.=$51^{\circ}$; systemic velocity V$_{sys}=3032$ km s$^{-1}$; maximum rotation velocity V$_{max}=210$ km s$^{-1}$ and radius at which maximum rotation velocity is attained R$_{max}=52''$ (10.9 kpc). ii) We estimated a dynamical mass inside 52$''$, based on the maximum rotation velocity to be M$_{dyn}=2\times10^{11}$ M$_{\odot}$. iii) We have detected a large velocity dispersions (up to 120 km s$^{-1}$) in the central region; from the bright regions in the west at about 10 kpc from the nucleus (with velocities up to 60 km s$^{-1}$) and from the bright regions just east of the end of the stellar bar (with velocities up to 70 km s$^{-1}$). The velocity dispersion from the central region might be associated with the presence of an outflow either from a compact nucleus or from a circumnuclear structure and regions of intense star formation with WR stars. iv) We observed deviations from the isovelocity contours from the south west (approaching side) outer most regions of the galaxy coinciding with the string of bright H$\alpha$ sources. A possible interpretation is that these sources are part of a semi-ring with a higher radial velocity, from 10 to 40 km s$^{-1}$, than expected at their radii assuming circular orbits, or that this region is at different inclination than the plane of the galaxy.

\section*{Acknowledgements}

	We would like to thank an anonymous referee for useful comments and suggestions on how to improve the paper. We acknowledge fruitful conversations with E. Moreno and V. Avila-Reese. M. Rosado acknowledges partial financial support from CONACYT (M\'exico) grant 27984-E and from DGAP.A. (UNAM, M\'exico) grant IN122298.

\normalsize

\clearpage
\newpage

\small 
\begin{table}
\caption[ ]{
Inclination of the galaxy NGC~3367 with respect to plane of sky
}
\begin{flushleft}
\begin{tabular}{lll}
\hline
$i$ &  Method & Reference \cr
\hline

37$^{\circ}$ & Axial ratio \& spiral structure & Danver (1942)\cr

32$^{\circ}$ & Axial ratio (RC2)   & de Vaucouleurs (1964)\cr

0$^{\circ}$ & Axial ratio (UGC)    & Nilson (1973)\cr

17$^{\circ}$ & Axial ratio (UGC)   & Williams \& Kerr (1981)\cr

25$^{\circ}$ & Fits to HI spectra              & Helou et al. (1981)\cr

25$^{\circ}$ & Axial ratio (RC2)   & Bottinelli et al. (1982)\cr

24$^{\circ}$ & Axial ratio (RC2)   & Huchtmeier \& Seiradakis (1985)\cr

6$^{\circ}$  & Fits to optical brightness distribution & Gr\"osbol (1985)\cr

19$^{\circ}$  & Axial ratio (UGC) & Tully (1988)\cr

30$^{\circ}$ & Axial ratio (RC3)   & de Vaucouleurs (1991)\cr

30$^{\circ}$ & Axial ratio (RC3)   & Ho et al. (1997)$^a$\cr

30$^{\circ}$ & Fabry Perot H$\alpha$   & This work \cr

\hline
\end{tabular}
\end{flushleft}
$^a$ Ho et al. (1997) reports in their table 11, column 10 values of inclination of disk to the line of sight, however, we believe that they probably meant inclination to the plane of the sky.
\end{table}

\small 
\begin{table}
\caption[ ]{
Parameters of Rotation Curve of the galaxy NGC~3367 in this work
}
\begin{flushleft}
\begin{tabular}{llllll}
\hline

$i$ &  P.A. ($^{\circ}$) & V$_{sys}$ (km s$^{-1}$) & V$_{max}$ (km s$^{-1}$) & R$_{max}('')$ & Parameter fixed\cr
\hline
30$^{\circ}$ & 52 & 3034 & 208   & 49 & none\cr

27$^{\circ}$ & 49 & 3028 & 241   & 37 & x,y,i receding only\cr

27$^{\circ}$ & 53 & 3037 & 291   & 230 & x,y,i approaching only\cr

27$^{\circ}$ & 52 & 3032 & 227   & 48 & x,y,i both sides\cr

31$^{\circ}$ & 51 & 3032 & 199   & 46 & x,y,P.A. both sides\cr

20$^{\circ}$ & 51 & 3032 & 302   & 55 & x,y,P.A.,$i$ both sides \cr

30$^{\circ}$ & 51 & 3032 & 204   & 48 & x,y,P.A.,V$_{sys}$ both sides\cr

25$^{\circ}$ & 46 & 3032 & 244   & 49 & x,y,P.A.,V$_{sys}$ both sides\cr

29$^{\circ}$ & 53 & 3032 & 213   & 51 & x,y,V$_{sys}$ both sides R$_i=5''$,R$_o=75''$\cr

30$^{\circ}$ & 51 & 3032 & 211   & 38 & x,y,P.A.,V$_{sys}$ receding side only $\pm60^{\circ}$ from P.A. \cr

30$^{\circ}$ & 51 & 3032 & 253   & 185 & x,y,P.A.,V$_{sys}$ approaching side only $\pm60^{\circ}$ from P.A. \cr

30$^{\circ}$ & 51 & 3032 & 210   & 52 & x,y,P.A.,$i$,V$_{sys}$ both sides only \cr

\hline
\end{tabular}
\end{flushleft}
\end{table}

\clearpage

\end{document}